\documentclass[fleqn,10pt]{wlscirep}
\usepackage[utf8]{inputenc}
\usepackage[T1]{fontenc}
\usepackage{lineno}
\usepackage[utf8]{inputenc}
\DeclareMathOperator{\sinc}{sinc}

\title{High-contrast, speckle-free, true 3D holography via binary CGH optimization}

\author[ ]{Byounghyo Lee}
\author[ ]{Dongyeon Kim}
\author[ ]{Seungjae Lee}
\author[ ]{Chun Chen}
\author[*]{Byoungho Lee}

\affil[ ]{School of Electrical and Computer Engineering, Seoul National University, Gwanak-Gu Gwanakro 1, Seoul 08826, South Korea}
\affil[*]{byoungho@snu.ac.kr}

\begin{linenumbers}
\begin{abstract}
Holography is a promising approach to implement the three-dimensional (3D) projection beyond the present two-dimensional technology. True 3D holography requires abilities of arbitrary 3D volume projection with high-axial resolution and independent control of all 3D voxels. However, it has been challenging to implement the true 3D holography with high-reconstruction quality due to the speckle. Here, we propose the practical solution to realize speckle-free, high-contrast, true 3D holography by combining random-phase, temporal multiplexing, binary holography, and binary optimization. We adopt the random phase for the true 3D implementation to achieve the maximum axial resolution with fully independent control of the 3D voxels. We develop the high-performance binary hologram optimization framework to minimize the binary quantization noise, which provides accurate and high-contrast reconstructions for 2D as well as 3D cases. Utilizing the fast operation of binary modulation, the full-color high-framerate holographic video projection is realized while the speckle noise of random phase is overcome by temporal multiplexing. Our high-quality true 3D holography is experimentally verified by projecting multiple arbitrary dense images simultaneously. The proposed method can be adopted in various applications of holography, where we show additional demonstration that realistic true 3D hologram in VR and AR near-eye displays. The realization will open a new path towards the next generation of holography.
\end{abstract}
\end{linenumbers}
\begin{document}

\flushbottom
\maketitle

\thispagestyle{empty}
\section*{Introduction}

Holography has been investigated with lots of attention based on the possibility of three-dimensional (3D) projection in the past decades\cite{poon2006digital,matsushima2005computer,yu2017ultrahigh,chlipala2019color}. True 3D holography requires ability to project any 3D volume with high-axial resolution and can control all 3D points independently. Recent emerging studies of computer-generated holograms (CGHs) concentrate on visualization of photorealistic scene, including high-contrast and speckle-free reconstruction\cite{shi2021towards,peng2020neural}. Yet, the works strict the 3D points following the specific smooth phase distribution to avoid speckle, which leads to dependency among the complex-valued 3D points. As disadvantages of the dependency, the 3D reconstruction of smooth phase encoding suffers low axial resolution with a small eyebox and difficulty in supporting parallax.

Although holography can reproduce the arbitrary 3D volume from the linear superposition of Fresnel zone plate patterns, if the phase of 3D points is defined to some specific distribution, it is limited to control the complex-valued holographic 3D points independently. A possible approach for breaking the dependency is to make the phase profile of 3D points following the random distribution so that the mutual interference is removed by using the orthogonality of random vectors\cite{makey2019breaking}. However, the speckle noise from the random phase severely hampers reconstructed images quality, so that  the previous works on true 3D projection are limited to sparse images such as dots or letters, or noisy images\cite{makey2019breaking,zhang20173d}. Until today, it remains as an unsolved challenge for holography to achieve high-contrast, speckle-free, and true 3D projection simultaneously\cite{chang2020toward}.

The speckle noise in holography appears inevitably due to the technical issue of the spatial light modulator (SLM), imaging principle of holography, and statistics of random phasors. The finite window size and limited diffraction angle of the SLM cause a single voxel to be imaged with area rather than an ideal point. There have to exist interference between the adjacent voxels, resulting in significant speckle contrast even if a small number of random phasors are summated\cite{goodman2007speckle}. Due to the  fundamental limitation, it is challenging to remove the speckle for random phase holography. Decreasing coherency of the light source (e.g., LED) is one available approach to mitigate the speckle, but it sacrifices overall depth range and spatial resolution\cite{kozacki2016color,lee2020light}. Another method is sampling reconstructed image considering the size of voxel's area to avoid interference between adjacent voxels, which suffers the loss of spatial resolution and the light efficiency\cite{mori2014speckle,takaki2011speckle}. Smooth phase methods reduce speckle by enforcing the interference between the adjacent voxels to constructive only \cite{chakravarthula2019wirtinger,choi2021optimizing,peng2020neural,shi2021towards,an2020slim,maimone2017holographic,chang2019computational}, but the dependency of phase results in the loss of the true 3D reconstruction capability.

The reconstruction quality and the ability of true 3D reconstruction are on the trade-off due to the speckle, which is hard to be resolved by a single CGH no matter how optimize it. To overcome the fundamental limitation, we extend the problem to finding a set of CGHs to give additional physical degree-of-freedom in the time-domain. From the statistical optics, one of the powerful strategy is to sum the mutually independent speckle patterns, known as temporal multiplexing (TM) in holography\cite{goodman2007speckle,lee2020wide}. However, implementation of speckle-free, full-color, and dynamic video projections through this method requires an SLM with a significantly fast refresh rate. The potential candidates are to utilize binary SLMs such as Ferro-electronic liquid crystal on silicon (FLCOS) and digital micromirror devices. However, the results from the binary SLMs inherently suffer from binary quantization noise\cite{pandey2011quantization}, appearing as background noise which reduces reconstructed image contrast \cite{buckley2011computer,buckley2011holographic}. The previous works for mitigating the binary noise are implemented through direct binary search\cite{liu2019enhanced, seldowitz1987synthesis}, error diffusion\cite{yang2019error,buckley2011real}, and Gerchberg–Saxton (GS) algorithm\cite{masuda2016improvement}. But the direct binary search is computationally heavy and the improvements by other methods are not obvious.

In this paper, we report an unprecedented realization of speckle-free, high-contrast, true 3D holography supporting dynamic video through the development of the binary optimization framework. The trade-off that cannot be physically solved in one hologram is overcome with multiple holograms, and the remaining \textit{reduce-able} binary noise is mitigated through a developed hologram optimization framework. We experimentally demonstrate the true 3D holography by multidepth projection of mutually independent images. The results show high-resolution, high-contrast, speckle-free, dense color images that go beyond state-of-the-art works that often represent a sparse and monochromatic imagery. We also apply high-quality true 3D holography to the near-eye displays as a demonstration that our approach can be adopted in various holographic applications. With the powerful performance of our true 3D holography, we verify the ability to provide realistic 3D hologram, showing high correspondence with rendered 3D scenes in virtual reality (VR) and real objects in augmented reality (AR).

\section*{Results}
\subsection*{Prototypes of true 3D holography}
To demonstrate the realization of true 3D holography, we build two types of full-color holographic prototypes, such as multiplane projection system and near-eye displays. The schemes of these setups are shown in Figs. \ref{fig:independent}a and \ref{fig:2d}a (see Supplementary Information for real photographs). We use an 8.2 µm pitch of FLCOS (Fourth Dimension Displays; QXGA-R10) as a binary SLM, synchronized with a fiber-coupled full-color laser diode (Wikioptics; Red: 638 nm, Green: 520 nm, Blue: 450 nm). The laser source is expanded and collimated by combining an objective lens (Nikon; M-40x, 0.65NA) and a collimating lens (Nikon; 50mm f/1.4). The FLCOS is originally designed to operate for 8 bits $\times$ 3 colors $\times$ 150 Hz when the resolution is 1920 $\times$ 1200 (WUXGA). We relocate the time sequence for the binary operation, which corresponds to the 1 bit $\times$ 3 colors $\times$ 50 Hz $\times$ 24 TM. The time sequence is synchronized with the fiber-coupled laser diode using the transistor-transistor logic (TTL) signal. Although our FLCOS is originally designed for the phase only modulation, we use it for the amplitude type by sandwiching two orthogonal linear polarizers to adopt single-side band encoding. The amplitude type CGH encoding makes our optimization focusing on minimizing the binary quantization noise only without considering the complex-valued representation, and enables successful optimization results. A custom-made single-sideband spatial filter is placed at the Fourier plane of the Fourier lens (Canon EF 200mm f/2.8). The filter size is determined to match the blue channel’s Fourier domain size of 11 mm x 5.5 mm. Since red and green channels have larger Fourier domain sizes than blue, input RGD data is resized and zero-padded to match the size of the blue channel. The size of the entire projection area of multiplane system is equal to the blue channel's Fourier domain size. For the near-eye setup, an additional eyepiece lens (Nikon; 50mm f/1.4) is placed by consisting 4f system with the Fourier lens. The prototype's field of view (FOV) is 12.6° $\times$ 6.3°, and the eyebox is 3.9 mm $\times$ 2.5 mm. 

\subsection*{Realization of speckle-free true 3D projection with arbitrary 3D volume}
\begin{figure}[ht]
\centering
\includegraphics[width=1\linewidth]{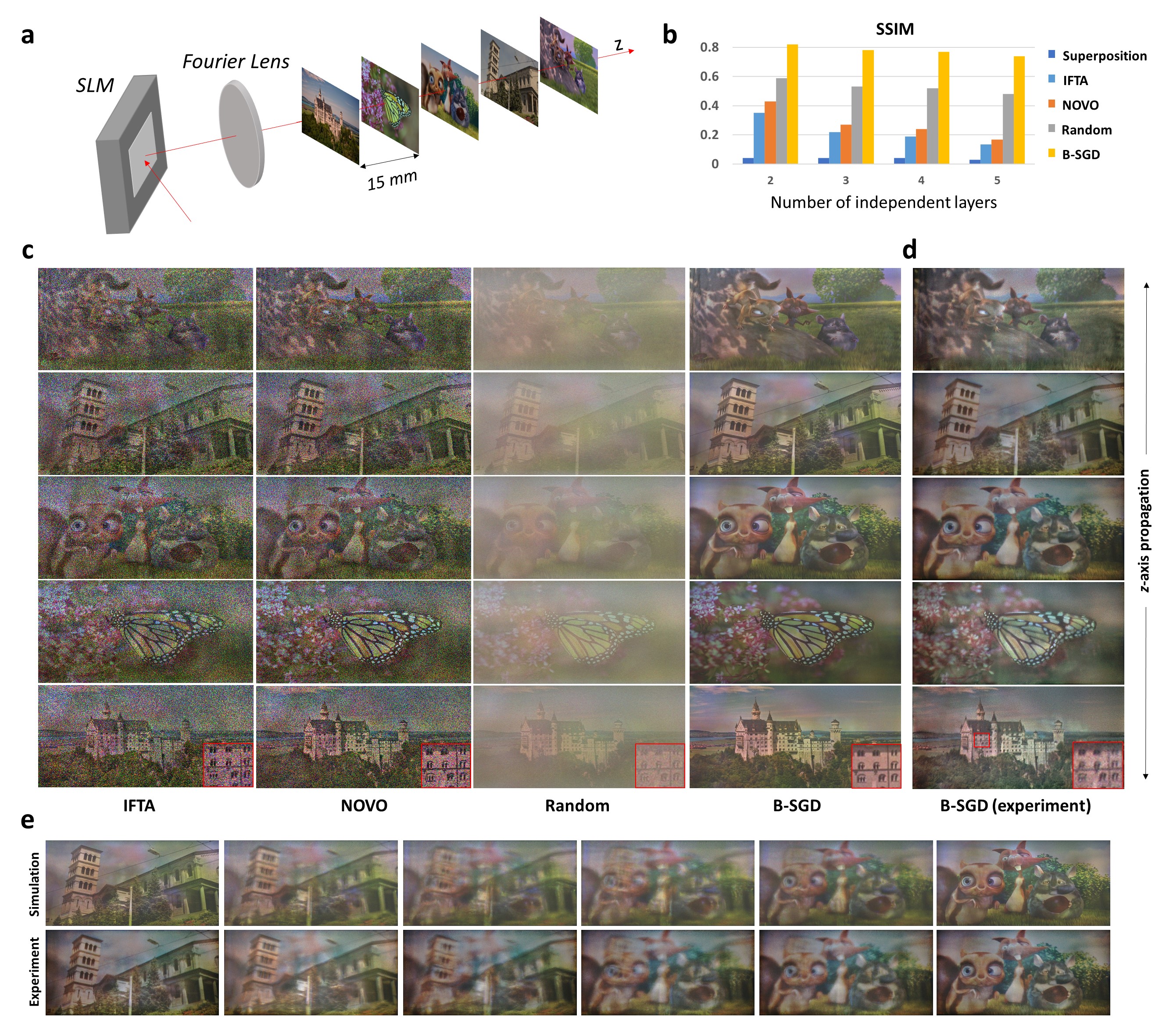}
\caption{\textbf{Demonstration and performance evaluation for crosstalk-free 3D projection.} \textbf{a.} Scheme of Fourier holographic projection, where five independent images are apart from 15 mm interval. \textbf{b.} Simulated comparison results of five CGH methods, including superposition within phase hologram (superposition), iterative Fourier transform algorithm (IFTA), non-convex optimization (NOVO), superposition within binary hologram (random), and binary-SGD (B-SGD). The binary holographic methods only adopt 24 TM considering dynamic projection. \textbf{c.} Numerical propagation results as focusing on each target depth layers. Only B-SGD shows crosstalk-free projection with high-contrast and low-speckle. \textbf{d.} Experiemental results of proposed B-SGD. \textbf{e.} Intermediate planes of B-SGD method. Source images by Kim, C. et al.\cite{kim2013scene}, Sheikh, H.R. et al.\cite{sheikh2006statistical}, and $\copyright$ copyright 2008, Blender Foundation | www.bigbuckbunny.org.}
\label{fig:independent}
\end{figure}

To demonstrate the ability of speckle-free true 3D holographic projection, we design multiplane projection systems whose target 3D volume is consisted of multiple dense images that are uncorrelated to each other. The five target intensities are spaced 15 mm apart centered on the Fourier plane as shown in Fig. \ref{fig:independent}a. Here, we select the distance by considering the depth of field of the holographic projection system, which is limited by the space-bandwidth product of the SLM. We compared our binary optimization method, termed binary-stochastic gradient descent (B-SGD), with the state-of-art works IFTA\cite{makey2019breaking} and NOVO\cite{zhang20173d}(see Methods for the procedure of B-SGD). We also compare with the superposition hologram generated by summation of propagated fields from all depths (termed Superposition) and its binary form combined with TM (termed Random). Since the size of Fourier plane differs by the hologram encoding methods such as phase-only or amplitude-only, we apply the anamorphic transform to the phase-only hologram to match the size\cite{kim2014anamorphic}. To compare the CGH methods quantitatively, we simulate the average structural similarity index measure (SSIM) according to the number of layers as shown in Fig. \ref{fig:independent}b. The simulated results for the case of the five independent layers are included in Fig. \ref{fig:independent}c. Since all methods use the random phase, it is available to project the independent images at each target depth with high-axial resolution. The IFTA and NOVO offer better results than Superposition and provide acceptable image quality when the target profiles are simple (see Supplementary Information), but the inevitable speckle noise from the random phase makes it challenge to be applied for the dense and complicated profiles as shown in Fig. \ref{fig:independent}c.

The binary CGHs such as Random and our B-SGD can overcome the speckle issue and provide higher SSIM than the above phase-only CGH results. The B-SGD optimization significantly improves image contrast compared to the Random case, and show the best performance of all the comparing methods. Figure \ref{fig:independent}d is experimental photographs of B-SGD hologram, confirming that the complicated color images can be reproduced with high-quality. Figure \ref{fig:independent}e is the intermediate planes between the third and fourth planes from the SLM, showing a high agreement between the experiments and simulations. An experimentally recorded axial sweep video within the full depth range is available in Supplementary Video 1. 

\begin{figure}[ht]
\centering
\includegraphics[width=0.8\linewidth]{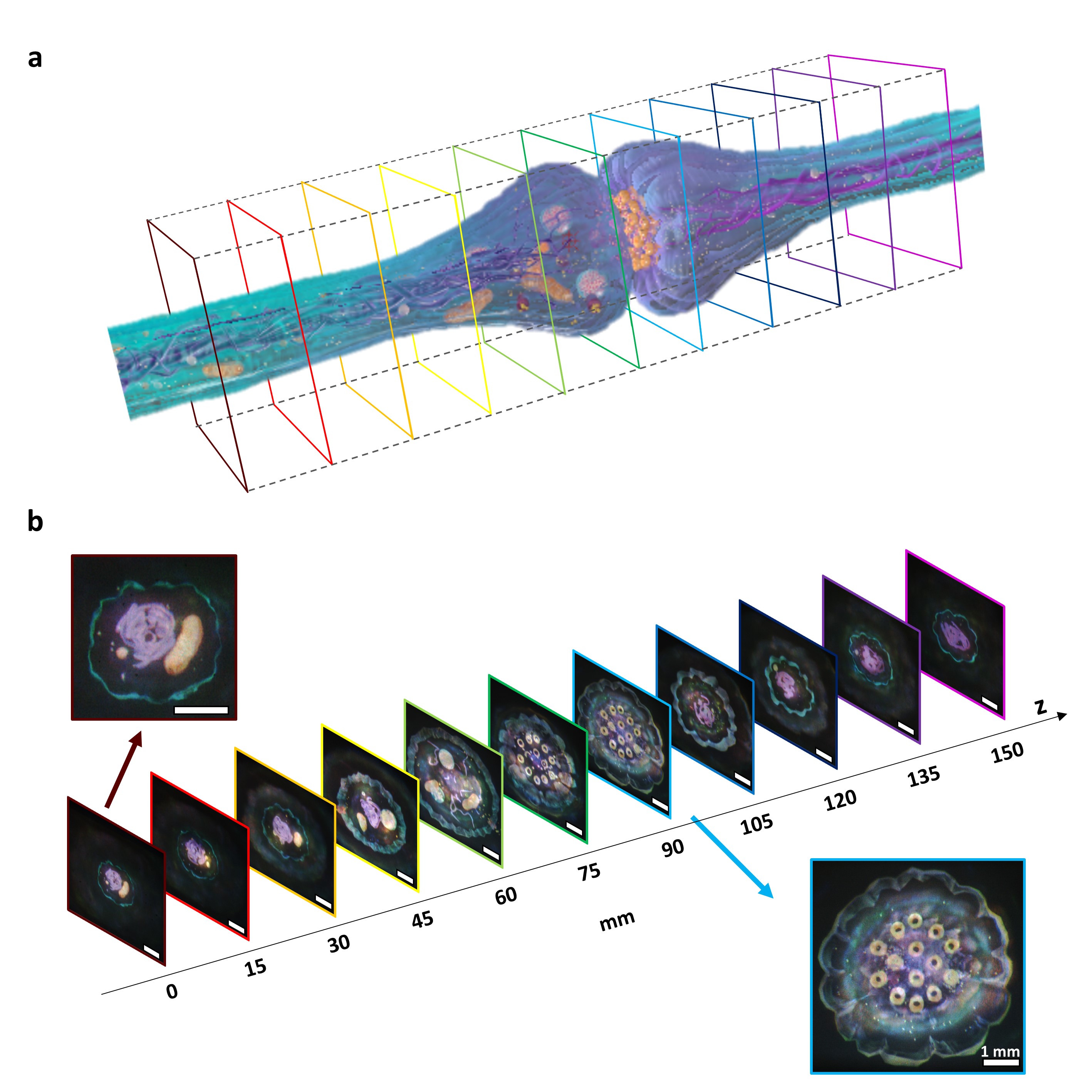}
\caption{\textbf{Tomographic 3D projection} \textbf{a.} Target 3D model ($Synapse$), purchased from Free 3D, All Rights Reserved. Eleven cross-section images are taken by using the 3D rendering software ('Autodesk Maya'). \textbf{b.} Experimental results of the tomographic projection. The results are captured by shifting the camera from 0 mm to 150 mm in the $z$-axis. White lines denote length of 1 mm.}
\label{fig:tomography}
\end{figure}

We design the tomographic 3D visualization as an application of the true 3D multiplane projection. Figure \ref{fig:tomography}a is the 3D blended model (\textit{Synapse}), where we obtain the eleven target tomographic images by capturing the cross-section of the model using the 3D rendering program. The target images are designed to be spaced apart by 15 mm for consistency with the above demonstration. Figure \ref{fig:tomography}b is experimentally captured images at each designed depth, showing the feasibility of tomographic visualization with high-quality and speckle-free projection. These results also support the possibility of controlling a large number of arbitrary depth planes in the wide depth range.

\subsection*{Evaluating the performance of true 3D holography}
\begin{figure}[ht]
\centering
\includegraphics[width=\linewidth]{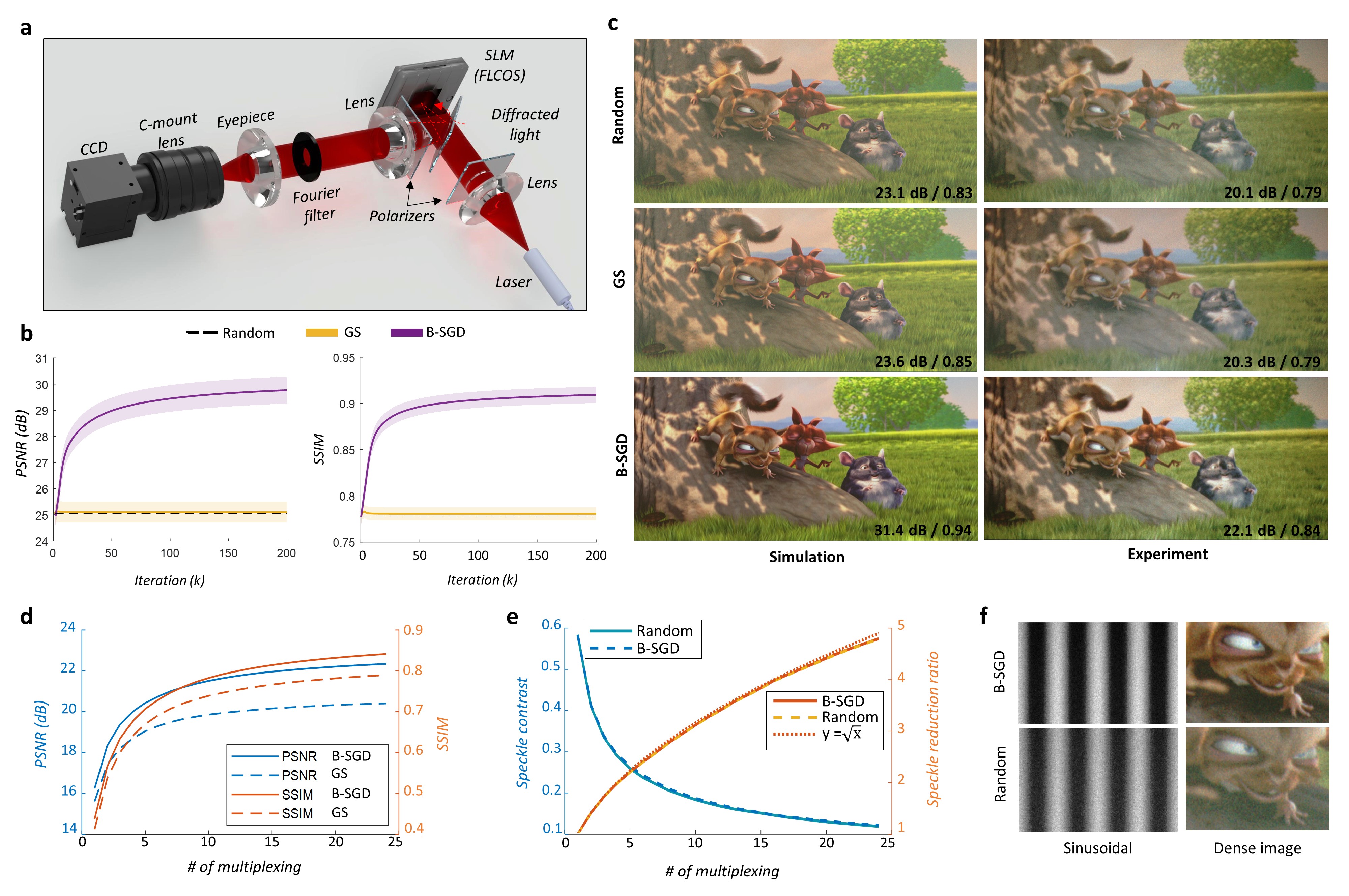}
\caption{\textbf{Demonstration and performance evaluation of true 3D holography in 2D near-eye displays.} \textbf{a.} Scheme of Fourier amplitude holographic display prototype. The figure visualizes SLM illuminating light and diffracted light from the center pixel of SLM (dot line). Only red wavelength is depicted for simplicity. \textbf{b.} Simulation results to assess three binary CGH methods, including superposition as updating random phase (random), Gearchberg-Saxton (GS), binary-SGD (B-SGD). \textbf{c.} (left) Simulation and (right) experimental results. All results are obtained with 24 TM. The bottom insets are PSNR and SSIM. \textbf{d.} Experimentally measured PSNR and SSIM according to number of multiplexing \textbf{e.} Experimentally measured speckle contrast and its reduction ratio. \textbf{f.} Experimental image contrast comparisons using the sinusoidal hologram and zoom-in region of (c). Source image by $\copyright$ copyright 2008, Blender Foundation | www.bigbuckbunny.org.}
\label{fig:2d}
\end{figure}

We simulate the holographic reconstructions to assess the performance of the B-SGD. Here, we use the near-eye display setup (see Fig. \ref{fig:2d}a) to reconstruct and evaluate the binary hologram. Figure \ref{fig:2d}b shows the simulated results corresponding to the iteration number using image evaluation metrics such as the peak signal-to-noise ratio (PSNR) and the SSIM. The GS method and non-optimized random phase results are compared by applying an equal number of TM (24 binary holograms). The Random results are illustrated by dot line because they are not achieved from iterative methods. The figures show mean values and standard error of 50 data sets from $\textit{Sintel}$\cite{sintel}, where the data sets are obtained by sampling five frames for each of the ten scenes ('market', 'cave', 'bamboo', etc.). The results confirm that the GS algorithm quickly falls into the local minima and the performance improvement is subtle compared to the initial and Random results. On the other hand, the B-SGD optimization shows a clear performance improvement in both PSNR and SSIM.

Next, we experimentally evaluate our system using a 2D image (from $\textit{bigbuck bunny}$\cite{Bigbuckbunny}) as shown in Fig. \ref{fig:2d}c. Although all the binary holographic methods are free from the speckle noise with the advantage of TM, the Random and GS suffer from background noise and low-image contrast. The proposed B-SGD only reconstructs the speckle-less and high-contrast holographic image in both experiment and simulation. The PSNR and SSIM are measured for quantitative evaluation, where these two metrics are denoted in bottom insets in Fig. \ref{fig:2d}c. The experimental results as increasing the number of multiplexing are summarized in Fig. \ref{fig:2d}d. Our prototype experimentally shows high performances that the PSNR of 22.1 dB and SSIM of 0.84. To our knowledge, the experimental values are highly comparable to the state-of-art results from smooth phase holography although we adopt random phase\cite{peng2020neural, choi2021optimizing}. Figure \ref{fig:2d}e shows the experimentally measured speckle contrast according to the number of multiplexing and its reduction ratio, where the target intensity has a flat constant. Interestingly, the speckle contrast shows the identical trend in both optimization or not, and the reduction ratios match well with the statistical theory. It is implied that our optimization preserves the uncorrelation among the speckle patterns, and the advantages of TM is achieved without sacrifice. We also measure the image contrast by displaying sinusoidal pattern hologram as shown in Fig. \ref{fig:2d}f. The Michelson contrast ${{({I_{\max }} - {I_{\min }})} \mathord{\left/
 {\vphantom {{({I_{\max }} - {I_{\min }})} {{\rm{(}}{{\rm{I}}_{\max }} + {I_{\min }})}}} \right.
 \kern-\nulldelimiterspace} {{\rm{(}}{{\rm{I}}_{\max }} + {I_{\min }})}}$ is improved from [0.83, 0.75, 0.73] to [0.91, 0.84, 0.82] for [R, G, B] color channels, where $I_{max}$ and $I_{min}$ are the maximum and the minimum intensities. See Supplementary Information for additional 2D experimental results.

\subsection*{Realization of realistic 3D holographic near-eye displays}
\begin{figure}[ht]
\centering
\includegraphics[width=0.8\linewidth]{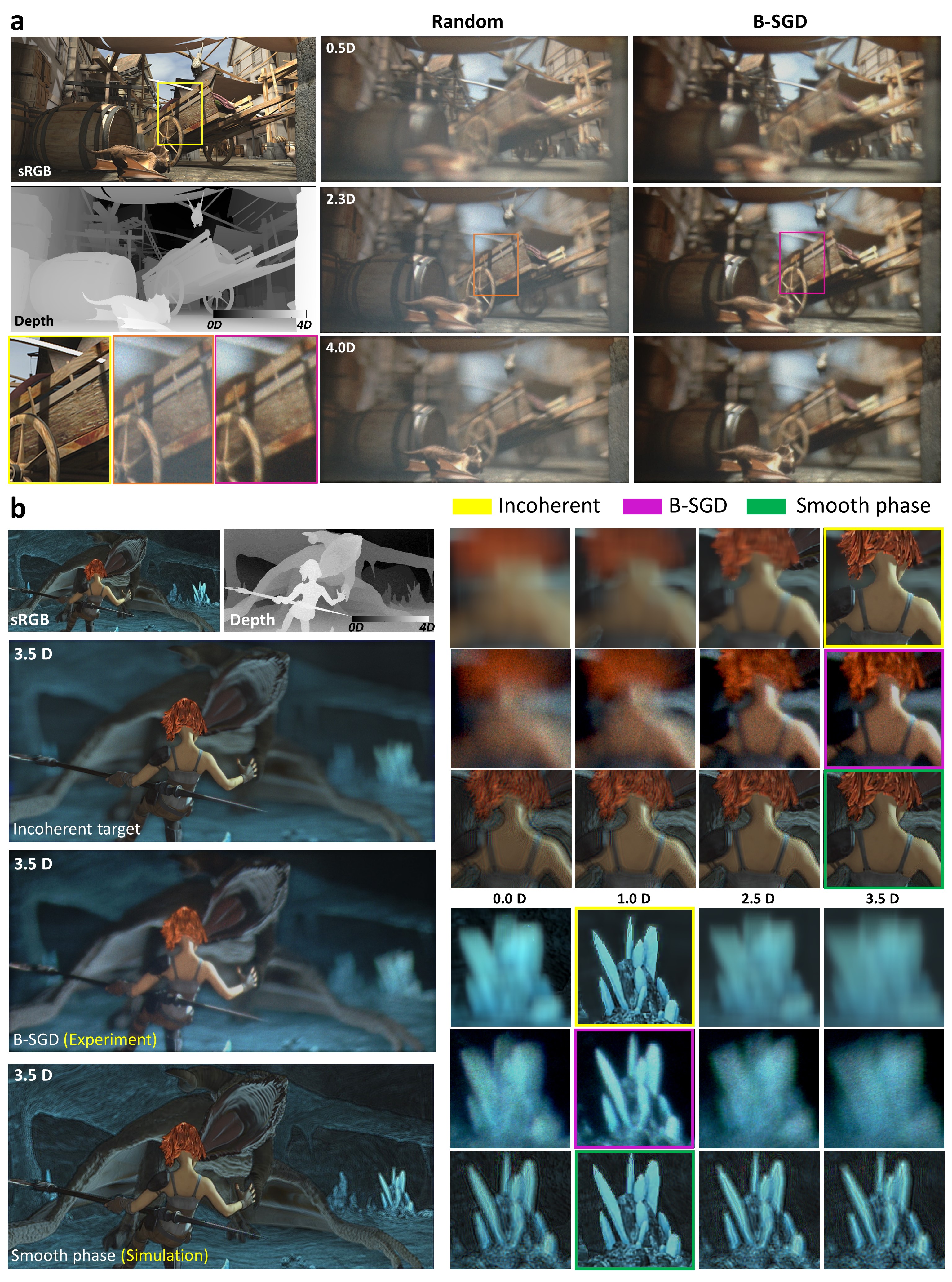}
\caption{\textbf{Experimental results for 3D scenes whose depth range extends between 0.0D to 4.0D.} \textbf{a.} Three example focal sweep scenes when the camera is focusing on 0.5D, 2.3D, 4.0D without/with optimization. \textbf{b.} Comparisons with incoherent 3D target and simulated smooth phase holography. Focal sweep results are summarized in right figures where focused images are noted as dot boxes. Our prototype shows realistic focus cues well match the 3D target. On the other hands, smooth phase fails to provide accurate focus cues. The source data by $\copyright$ copyright Blender Foundation | durian.blender.org.}
\label{fig:3-D}
\end{figure}

Since our realization overcomes the fundamental challenge for high-quality true 3D holography, it can be applied in various holographic applications. Here, we show an application to the holographic near-eye displays, and demonstrate the proposed method can provide not only speckle-free, high-contrast but also realistic focus cues and natural parallax. We use RGBD data as input and set the number of depth layers $N$ = 32, where the distances for nearest and farthest are 4.0 diopter (D) and 0.0D. The depth distortion from chromatic aberration is experimentally corrected by arranging depth range and each color's FOV. Figure \ref{fig:3-D}a shows some examples of the experimentally captured focal images where the top insets denote the focal depth of the camera. In the same manner of the 2D case, the random binary holography supports speckle-free imagery but its image contrast is low due to the binarization noise. On the other hand, the proposed B-SGD achieves high-contrast results without sacrificing focus cues, occlusion, spatial resolution. Since the random phase fulfills the entire eyebox and provides the highest axial resolution, significant 3D focus and defocus effects are observed. Fig. \ref{fig:3-D}b shows comparisons of the shape of focus and blur among the target, our experimental results, and simulated smooth phase holography. Note that the target is calculated from the incoherent transfer function to reflect the 3D focus cues of real world (see Methods). The experimental photographs show that the proposed B-SGD successfully projects 3D space, including focus and defocus scenes, matching the incoherent real world at all depths. To emphasize the significance of our work, we simulate smooth phase holography that encodes phase of target planes follows smooth distribution instead of random. Here, we simulate the ideal case where the complex amplitude is perfectly represented by amplitude encoding and the hologram is reconstructed through perfect propagation model (see Discussion). Although the ideal simulation shows speckle-free results at the focused plane, it fails to provide natural and realistic focus cues compared to the incoherent 3D target. More results of continuous focal-sweep video is available from Supplementary Video 2. The full-color synchronized 50 Hz video is available from Supplementary Video 3, which supports the capability of dynamic projection of our system.

\begin{figure}[ht]
\centering
\includegraphics[width=1\linewidth]{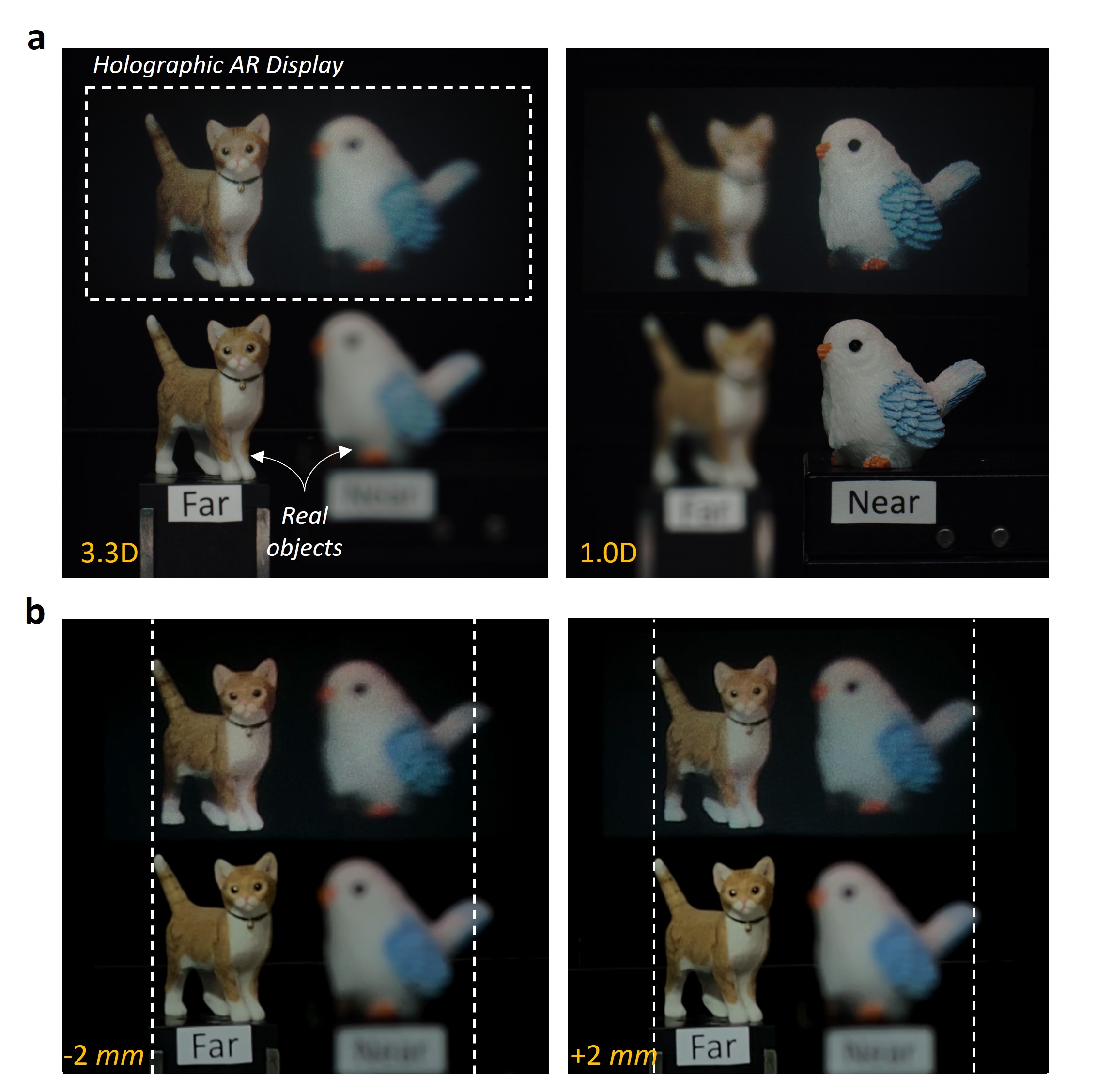}
\caption{\textbf{AR demonstration of true 3D holography featuring realistic focus cues and parallax} The real objects  are placed on the 3.3D (bird) and 1.0D (cat). \textbf{a.} Focal sweep results of our AR prototype, where the hologram shows realistic focus cues that of the real object. \textbf{b.} Parallax results by laterally shifting the camera. Bottom insets are the shifted distance of the camera, and dot lines indicate the identical FOV in both results. The hologram and real objects are captured simultaneously.}
\label{fig:ar}
\end{figure}

Finally, beyond the comparison with the computed virtual 3D scene, we compare with the real objects to show the implementation of the most realistic 3D holographic projection. We convert our VR prototype to the AR type by placing an additional beam splitter between the eyepiece lens and camera lens (see Supplementary Information). Two real objects are placed on the 3.3D (bird) and 1.0D (cat), and the target images are achieved using the built AR prototype. Here, we capture two target images focusing on each object and digitally synthesize the target RGBD data. The aperture stop of the C-mount lens is set at about 3.3 mm, considering that the eyebox of our holographic system is 3.9 mm $\times$ 2.5 mm. Figure \ref{fig:ar}a is the AR result as changing the camera focal state to each object. The result confirms that our true 3D holography provides natural and realistic focus cues that correspond to real objects. In addition, since our system truly reconstructs the 3D objects, realistic parallax is also provided, as shown in Fig. \ref{fig:ar}b. Here, the parallax results are captured by a mobile phone whose aperture diameter is 3 mm, where the white dot lines represent identical FOV. More results for the focal-sweep video and the parallax video of the AR experiments are available in Supplementary Videos 4 and 5. 

\section*{Discussion}
Here, we briefly discuss the importance of the random phase distribution in 3D holography through comparison with the smooth phase (SP) holography. Recently, variety of works such as double-phase \cite{maimone2017holographic, shi2021towards}, Wirtinger holography \cite{chakravarthula2019wirtinger}, and SGD optimization \cite{peng2020neural} encode the CGH following the smooth distribution to reduce speckle. The methods have a common goal to represent the complex amplitude using phase-only CGH. Rather than comparing with all the cases, we select amplitude hologram with SP as a representative method since it can completely express the complex amplitude. To avoid any difference from Fourier or Fresnel hologram type, we simulate the SP holography in the Fourier type. While our random phase-based method shows a natural focal blur suitable for each depth, SP holography provides a large depth-of-field and almost all-in-focused 3D scenes(see Supplementary Information). In addition, the shape of the blur for SP is a diffraction pattern determined by the spatial frequency of the amplitude. It is inconsistent with the real world, where the point spread function is not correlated with spatial frequency of the texture. Also, almost energy is concentrated at the DC for SP holography, so the effective Fourier spectrum is much smaller than the SLM supporting diffraction region. If the observer's eye pupil deviates from the DC, the entire image cannot be observed, and only the high-frequency component of amplitude (edge) is perceived. Thus, parallax cannot be provided by SP holography. 

The reproducibility of the speckle-free holographic displays is an important issue due to the sensitiveness of the coherent imaging. Peng et al. and Chakravarthula et al. deal with the speckle noise generated in the mismatch between the ideal and the actual wave propagation and summarizing the difficulty in reproduction for several SP holographic displays \cite{peng2020neural,chakravarthula2020learned}. The reproducibility depends on the strategy chosen to alleviate the speckle noise. SP holography reduces the speckle by equalizing the phase between adjacent reconstructed points so that only constructive interference occurs. However, it is susceptible to experimentally generated phase disturbing noise, which leads to unwanted destructive interference. Thus, accurate phase compensation (sub-wavelength level) is required like the camera-in-the-loop technique, which also has a limitation to be repeated when the optical system, camera, or target image suffers any changes. On the other side, we adopt statistical reduction by averaging the uncorrelated speckle patterns through the time domain. Even if the phase disturbing random noise is added, each reconstructed point still follows the random distribution, and the speckle is removed. Hence we do not need any additional object phase compensation, and we achieve the successful speckle reduction in every experimental reconstruction. We believe our speckle reduction strategy is a more suitable approach considering the commercialization of holography in the next generation.

Recently, dynamic computer-generated holography (DCGH) which seems similar to our works is proposed \cite{curtis2021dcgh}. The main difference is that our B-SGD optimizes a binary hologram at once and repeats the whole process $m$ times for TM, but DCGH optimizes all the $m$ holograms at once considering dependency among the binary holograms. From the statistical speckle theory, the successful speckle reduction that reaches to the ratio of 1/$\sqrt{m}$ is achieved when the reconstructed intensity follows identical distribution. If each intensity is designed to follow some different distribution, then the speckle reduction ratio would be reduced. Thus, we believe that our approach that repeats frame-wise optimization to make every individual intensity profile following identical distribution is more appropriate to apply the TM for speckle reduction.

There are some issues for true 3D holography, which will be valuable topics for future research. First, since we take TM to reduce the speckle noise, a proportionally increased computational burden is an remaining issue for real-time implementation. We believe it will be resolved in the near future with the rapidly developing computational holographic works such as complex amplitude optimization \cite{chen2021multi}, and the deep-learning hologram generation \cite{horisaki2018deep, lee2020deep, peng2020neural, shi2021towards, eybposh2020deepcgh}. Note that the computing cost for optimizing a single B-SGD hologram has almost similar cost for a conventional single 8-bit phase-only SGD hologram. Second, the limited space-bandwidth product is a remaining challenge. Several space-bandwidth product extending approaches such as beam steering with eye-tracking \cite{jang2018holographic}, structured illumination with TM \cite{lee2020wide}, and placing photon sieve \cite{park2019ultrathin, kuo2020high} would be solutions to mitigate the issue. Third, an RGBD image shows a 3D scene from a viewpoint, so it does not have information about the object occluded at the current viewpoint. Although we start from a single RGBD image considering versatility and data bandwidth for near-eye displays, it is not necessary as we demonstrate in independent multi-plane projections. Creating the accurate target focal stack from the set of RGBD images or light-field data set taken from multiple viewpoints can solve the issue. Finally, the ability to implement the high-contrast, speckle-free, true 3D holography benefits not only for holographic displays in VR and AR but also for hologram-based 3D printing, lithography, and metasurfaces. We believe that our research will be a turning point in holography for practical utilization in various applications beyond the current academic research.

\section*{Methods}
\subsection*{Space conversion from sRGB to linear}
Since holography reconstructs light without any gamma correction, preprocessing for target intensity is required for the input sRGB images to be perceived correctly by humans. The sRGB to linear conversion relationship is given by 

\begin{equation}
    \begin{array}{l}
{I_{lin}} = \frac{1}{{12.92}}{I_{sRGB}}{\;\;\;\;\;\;\;\;\;\;\;\;\;\;\;\;\;\;\;\;\;\;\;\;\;\;\;\;\;\:}{\rm{       0}} \le {{\rm{I}}_{sRGB}} \le 0.04045\\\\
{I_{lin}} = {\left( {\frac{{{I_{sRGB}} + 0.055}}{{1.055}}} \right)^{2.4}.}{\rm{  \;\;\;\;\;\;\;\;\;\;\;\;\;\;\;\;\;\;}}0.04045 < {{\rm{I}}_{sRGB}} \le 1
\end{array}
\end{equation}
In our CGH calculation procedure, we first convert the input sRGB intensity image to the linear space intensity and sequentially achieve the linear space amplitude $A_{lin} = \sqrt{I_{lin}} \approx{} (I_{sRGB})^{1.1}$. The similarity between amplitude in linear space and intensity in sRGB space guarantees optimal perceived image for human observers by optimizing the amplitude in linear space\cite{peng2020neural}.
 
\subsection*{Generating target focal scenes for true 3D holography}
Our optimization is designed to find optimal binary hologram with minimal amplitude loss compared to the target 3D amplitudes. Thus, it is crucial to generate an optically accurate 3D target for realistic and crosstalk-free implementations. In this paper, we demonstrate two types of holographic systems that project an RGBD data and multiple independent images. From the RGBD data, we generate a target focal stack from spatially incoherent wave propagation to reflect the focus cues in nature and achieve a realistic 3D scene. The incoherently propagated intensity is given by 

\begin{equation}
{{\tilde T}_z}\left[ I \right]{\rm{ }} = \iint{\rm{ }}{\textit{F}}\left[ {I(x,y)} \right]{\rm{ }}{{H}_i}({f_x},{f_y}){\rm{ }}{e^{i2\pi ({f_x}x + {f_y}y)z}}d{f_x}d{f_y}{\rm{  }},
\label{eq:incoherent_transfer}
\end{equation}
where \textit{F} is Fourier transform, $f_x$, $f_y$ are spatial frequencies, $z$ is propagation distance, and ${{H}_i}$ is the incoherent free-space propagation transfer function. Using the relationship between incoherent and coherent transfer functions, ${{H}_i}$ can be described by 

\begin{equation}
\begin{array}{l}
{H_i} = {H_c} \star {H_c}\\
{H_c}({f_x},{f_y}) = \left\{ {\begin{array}{*{20}{c}}
{{e^{i\frac{{2\pi }}{\lambda }\sqrt {1 - {{\left( {\lambda {f_x}} \right)}^2} - {{\left( {\lambda {f_y}} \right)}^2}} z}},{\rm{\;\;\;\;\;\;\;\;  }}\sqrt {{f_x}^2 + {f_y}^2}  < \frac{1}{\lambda },}\\
{0{\rm{\;\;\;\;\;\;\;\;\;\;\;\;\;\;\;\;\;\;\;\;\;\;\;\;\;\;\;\;\;\;\;\;\;\;\;\;\;\;\;\;\;\; otherwise}}}
\end{array}} \right.
\end{array}
\end{equation}
where $\star$ denotes auto-correlation, $\lambda$ is wavelength, and ${{H}_c}$ is the coherent transfer function that widely used for angular-spectrum method\cite{goodman2005introduction}. Then, the target focal 3D scene is calculated by sum of the incoherently transferred intensities at all depths with consideration of occlusion:

\begin{equation}
I_n^T = \sum\limits_{k \in N} {{{{\tilde T}}_{\Delta {z_{k,n}}}}\left[ {{I_{x \in {x_k}}}} \right]{M_{k,n}}.{\rm{  \;\;\;\;\;\;\;\;\;\;   }}(for{\rm{ \;\; }}RGBD)} 
\end{equation}
Here, $n$ is index of depth layer ($n = 1,2,\dots,N$), $I$ is a given RGB image, ${x}_k$ is the set of object points at the $k$-th depth layer, $\Delta {z_{k,n}} = |z_{n} - z_{k}|$, and ${M_{k,z}}$ is the occlusion mask.

For the second type of input data (multiple independent images $I_{n}$), the target set directly corresponds to the input images. One additional step is to match the energies of all depth images according to each color channel to satisfy the energy conversation principle. To match the energies of all depth layers, we divide the intensity at each depth by its energy (the sum of the intensities for all $x$) and multiply with the average of all the layer energies. The target intensities are given by 

\begin{equation}
    I_n^T = \frac{1}{N}\left( {\frac{{\sum\limits_{n \in N} {\sum\limits_{\forall x} {{I_n}} } }}{{\sum\limits_{\forall x} {{I_n}} }}} \right){I_n}.{{\rm{  \;\;\;\;\;\;\;\;\;\;   }}(for{\rm{ \;\; }}Multiplane)} 
\end{equation}

\subsection*{Binary CGHs optimization}
The summarized binary CGH optimization procedure is visualized in Supplementary Information. We first initialize binary hologram with a designed holographic reconstruction system. The Fourier type holography is adopted where a single object point is reconstructed by the interference of most pixels of the SLM, which guarantees more possibility to find optimal hologram. The initial object phase is set by $2\pi$ range of random phase for maximum axial resolution. From the input data and the initial random phase, the initial complex amplitude at the SLM plane is achieved through coherent transfer function $H_{c}$ and Fourier transform $\textit{F}$. Next, an amplitude type of Fourier hologram ${H_{amp}}$ is generated from the complex amplitude using the single-sideband encoding method, which represents the complex amplitude using an amplitude-only SLM\cite{wyrowski1990diffraction}. It is converted to the initial binary hologram ${H_{bi}}$ by the forward binary operator, where we use the hard-clipping method ${B_{forward}}(x) = sign(x)$. Then, the focal amplitude stack for all depth layers ${A_{n}} (n \in N)$ are numerically reconstructed from the initial binary hologram.

After the initialization step, we calculate amplitude loss for all depth layers and summate those losses for the total 3D loss. The total volumetric loss enforces our optimization to minimize amplitude difference not only for focused but also for the blurred scenes in all depths. To achieve the optimized binary hologram, we set the problem as finding the optimum continuous amplitude hologram (${H_{amp}}$) resulting in minimum loss after binarization. This indirect problem setting allows for an incremental hologram update during gradient-based optimization. Our amplitude optimization can be summarized by solving the following optimization problem. 

\begin{equation}
\mathop {minimize}\limits_{{H_{amp}}} \sum\limits_{n \in N} {\sum\limits_{\forall x} {{\textit{L}}\left[ {{e_n}{A_n},\sqrt {I_n^T} } \right],} }
\label{problem}
\end{equation}
where $\textit{L}$ is the loss function and $e_{n}$ is an energy scale factor given by ${e_n} = \sqrt {{\textstyle{{\sum\limits_{\forall x} {I_n^T} } \over {\sum\limits_{\forall x} {{{\left| {{A_n}} \right|}^2}} }}}}.$ We set the $\textit{L}$ as the $l_{2}$ mean square error (MSE) considering that the MSE is sensitive with the difference of mean value rather than variance, which is appropriate to reduce the background noise of binary hologram. To solve the equation (\ref{problem}), We utilize the iterative optimization based on SGD designed for binary system, termed by B-SGD. Our update rules in iteration $k$ $(k = 1, 2, \dots, K)$ with learning rate $\alpha$ are 

\begin{equation}
H_{amp}^k = H_{amp}^{k - 1} - \alpha {\left( {\frac{{\partial {\textit{L}}}}{{\partial {H_{amp}}}}} \right)},    
\label{update}
\end{equation}
where $K$ is set as 200 in this paper. Unlike the previous phase-only SGD works, the backward gradient of the loss cannot be obtained directly because the binary operation during the forward propagation is non-diffentiable. Inspired by the binarized neural networks research\cite{courbariaux2016binarized}, we adopt the straight-through estimator, which assumes the backward binary operator using a similar shape of differentiable function. We choose the hard hyperbolic tangent function for ${B_{backward}(x)} = Htanh(x) = max(-1, min(1,x))$ as the estimated backward binary operator. After finishing the optimization process, the output optimized binary hologram is acquired through a forward binarization to the optimized amplitude hologram with the conversion of the -1 value to zero. The B-SGD approach has advantage for fast hologram computing with a graphic processing unit (GPU). Also, the gradient of the loss function in equation (\ref{update}) can be robustly achieved via autograd of Pytorch without the need to develop gradient manually. 

\subsection*{Speckle analysis for random phase holography}
We show the theoretical speckle analysis to ensure the fundamental challenge for random phase holography to be speckle-free. Here, we consider the Fourier holography system consisted of an SLM and a Fourier lens. The SLM has $n_x \times n_y$ pixels whose pitch correspond to $d_x \times d_y$ and the focal length of Fourier lens is $f$. At the Fourier plane, the intensity distribution of a reconstructed point followed $\sinc^2$ function due to the finite size of the SLM (or finite numerical aperture):

\begin{equation}
I = {I_0}\sinc{^2}\left( {\frac{{{n_x}{d_x}}}{{\lambda f}}x} \right)\sinc {^2}\left( {\frac{{{n_y}{d_y}}}{{\lambda f}}y} \right).
\end{equation}
By assuming that the main lobe only affects the interference with the adjacent points, the 2D size of the main lobe is given by 

\begin{equation}
  {A_{lobe}} = \frac{{4{\lambda ^2}{f^2}}}{{{n_x}{n_y}{d_x}{d_y}}}.
\end{equation}
From the sampling theorem, the physical interval between adjacent points is $\left( {{{f\lambda } \mathord{\left/
 {\vphantom {{f\lambda } {{n_x}{d_x},}}} \right.
 \kern-\nulldelimiterspace} {{n_x}{d_x},}}{\;}{{f\lambda } \mathord{\left/
 {\vphantom {{f\lambda } {{n_y}{d_y}}}} \right.
 \kern-\nulldelimiterspace} {{n_y}{d_y}}}} \right)$, and the 2D area allocated for each sampled point is calculated by 
 
 \begin{equation}
     A_{{\mathop{ interval}}} = \frac{{{\lambda ^2}{f^2}}}{{{n_x}{n_y}{d_x}{d_y}}}.
 \end{equation}
Then, the overlapped ratio by the $\sinc^2$ mainlobe corresponds to $r = \frac{{{A_{lobe}}}}{{{A_{interval}}}} = 4$, which is constant rather than function of the parameters. It implies that no matter how CGH is optimized and reproduced, there will be interference between adjacent pixels.

When the object phase is uniformly distributed on (-$\pi$,$\pi$), the speckle phenomena can be statistically analyzed through the number of overlapped phasors. We assume the length of phasors from adjacent points are identical to $a$ to simplify statistical formulas, where it is supported by the fact that neighboring pixels in general images have similar intensity. From the first-order statistical properties of the speckle\cite{goodman2007speckle}, the probability density function of intensity with the sum of finite number $r$ phasors is given by

\begin{equation}
    p(I) = 2{\pi ^2}\int\limits_0^{2\pi } {\rho J_0^r\left( {2\pi a\rho } \right)} {J_0}\left( {2\pi \sqrt I \rho } \right)d\rho,
\end{equation}
where $J_0$ is the Bessel function of the first kind. Then, we obtain speckle contrast $C$ by sequentially calculating expectation value and variance: 

\begin{equation}
\begin{array}{l}
E\left[ I \right] = {a^2},{\rm{\;\;\;\;    }}{\sigma ^2} = \left( {1 - \frac{1}{r}} \right){a^4},\\\\
{\rm{C}} = \frac{\sigma }{{E\left[ I \right]}} = \sqrt {1 - \frac{1}{r}}. 
\end{array}
\end{equation}
Note that the shape of speckle contrast increases rapidly from zero to one in the small number of $r$, and in our case with $r$ = 4, it becomes a high value of 0.87. Although many parameters affect the speckle contrast, including fill factor of SLM, intensity profile, and high-order lobes, the results sufficiently show the inevitable speckle noise from the random phase and justify the need of TM. Note that the analysis is derived in the linear space. 

\subsection*{Speckle reduction through TM}
When the object phase follows the random phase distribution, the random constructive and destructive interference among adjacent points results in inevitable speckle. One available approach for speckle reduction is to adopt the TM technique following the statistical speckle theory\cite{goodman2007speckle}, where the speckle variance is decreased by averaging the independent speckle patterns during the observing framerate. The independent speckle patterns are achieved by changing the initial object phase when calculating CGHs to follow other random phase sets. The accumulated intensity has reduced speckle contrast as ${C_m} = \frac{{{C_1}}}{{\sqrt m }}$, where $m$ $(m = 1, 2, \dots, M)$ is the number of TM. We choose the $M =24$ to avoid flickering observation by considering the human visual system. Using the off-the-shelf binary SLM which supports 3600 Hz in WUXGA resolution, we build the 50 Hz video projection prototypes in full-color. Here, the 50 Hz is given by dividing 3600 Hz into 3 colors and 24 TM. The speckle contrast is reduced by 4.9 times ($\sqrt{24}$) compared with the reconstructed scene of a single-frame random phase hologram. 

\subsection*{Camera acquisition method}
To capture the experimental results in this paper, we mainly use a machine vision CCD sensor (FLIR; GS3-U3-91S6C) adapted with a C-mount lens (Nikon; 60 mm f/2.8) for the multiplane projection and with another C-mount lens (Navitar; 35 mm f/1.4) for  near-eye VR system. The CCD sensor is controlled by the manufacturer providing GUI software (Spinview). Specifically, the gain and the sensor's black level are set to zero, and shutter speed is set to 20 ms. Sensor gamma is set to (1/2.2) to capture the sRGB intensities from the linear space intensities. Focal sweep videos (Supplementary Videos 1, 2, and 4) are recorded using the CCD camera. A mobile phone (Apple; iPhone pro 11) is used to achieve the high-framerate movie (Supplementary Video 3) and AR parallax results (Fig. \ref{fig:ar}b and Supplementary Video 5). Here, the telephoto lens (52 mm equivalent, f/2) is used whose actual focal length is 6 mm and the diameter of the aperture is 3 mm. The sRGB results are obtained automatically through the image processor built in the mobile phone. The white balance for all color results is manually matched by controlling the currency of the fiber-coupled laser diodes before capturing the experimental results.

\section*{Data availability}
The data to support the results within this paper is available from the corresponding author on reasonable request. 
\bibliography{sample}

\section*{Acknowledgements}
This work was supported by Institute for Information \& Communications Technology Promotion (IITP) grant funded by the Korea government (MSIT) (No. 2017-0-00787, Development of vision assistant HMD and contents for the legally blind and low visions).

\section*{Author contributions statement}
B.L. conceived the idea, realized the proposed method, built optical systems, conducted the experiment, and wrote the manuscript. D.K. executed the experiment for evaluation, analyzed the results, and wrote the manuscript with B.L. S.L. analyzed the results, and verified the optimization algorithm. C.C. performed and verified the optimization algorithm with the optical simulation. B.L. supervised the project. All of the authors discussed the results and reviewed the manuscript.

\section*{Additional information}
\textbf{Competing interests} The authors declare no competing interests. 

\section*{Supplementary information}

\end{document}